\documentclass{emulateapj}

\usepackage{natbib}

\usepackage{graphicx}
\usepackage{latexsym}
\usepackage{amsfonts,amsmath,amssymb}
\usepackage{url}
\usepackage[utf8]{inputenc}
\usepackage{fancyref}
\usepackage{hyperref}
\hypersetup{colorlinks=false,pdfborder={0 0 0},}

\newcommand{\ha}{\rm{H}\alpha}

\def\lya{Ly$\alpha$} 
\def\muv{$M_{\rm UV}$} 

\shorttitle{Probing the $z>6$ Universe with the HFF Cluster A2744}
\shortauthors{Hakim Atek}

\begin{document}


\title{Probing the $z>6$ Universe with the First Hubble Frontier Fields Cluster Abell 2744 \footnotemark[$\dagger$]}

\footnotetext[$\dagger$]{Based on observations made with the NASA/ESA Hubble Space Telescope, which is operated by the Association of Universities for Research in Astronomy, Inc., under NASA contract NAS 5-26555. These observations are associated with programs 13495 and 11689. Based in part on observations made with the Spitzer Space Telescope, which is operated by the Jet Propulsion Laboratory, California Institute of Technology under a contract with NASA. This work utilizes gravitational lensing models produced by PIs Ebeling, Merten \& Zitrin, and Sharon funded as part of the HST Frontier Fields program conducted by STScI. STScI is operated by the Association of Universities for Research in Astronomy, Inc. under NASA contract NAS 5-26555. The lens models were obtained from the Mikulski Archive for Space Telescopes (MAST).}

\author{Hakim Atek\altaffilmark{1}}
\author{Johan Richard\altaffilmark{2}}
\author{Jean-Paul Kneib\altaffilmark{1,3}}
\author{Benjamin Clement\altaffilmark{4}}
\author{Eiichi Egami\altaffilmark{4}}
\author{Harald Ebeling\altaffilmark{5}}
\author{Mathilde Jauzac\altaffilmark{6}}
\author{Eric Jullo\altaffilmark{3}}
\author{Nicolas Laporte\altaffilmark{7,8}}
\author{Marceau Limousin\altaffilmark{3}}
\author{Priyamvada Natarajan\altaffilmark{9}}

\altaffiltext{1}{Laboratoire d'Astrophysique, Ecole Polytechnique F\'ed\'erale de Lausanne, Observatoire de Sauverny, CH-1290 Versoix, Switzerland}
\altaffiltext{2}{CRAL, Observatoire de Lyon, Universit\'e Lyon 1, 9 Avenue Ch. Andr\'e, 69561 Saint Genis Laval Cedex, France}
\altaffiltext{3}{Aix Marseille Universit\'e, CNRS, LAM (Laboratoire d'Astrophysique de Marseille) UMR 7326, 13388, Marseille, France}
\altaffiltext{4}{Steward Observatory, University of Arizona, 933 North Cherry Avenue, Tucson, AZ, 85721, USA}
\altaffiltext{5}{Institute for Astronomy, University of Hawaii, 2680 Woodlawn Drive, Honolulu, Hawaii 96822, USA}
\altaffiltext{6}{Astrophysics and Cosmology Research Unit, School of Mathematical Sciences, University of KwaZulu-Natal, Durban, 4041 South Africa}
\altaffiltext{7}{Instituto de Astrofisica de Canarias (IAC), E-38200 La Laguna, Tenerife, Spain}
\altaffiltext{8}{Departamento de Astrofisica, Universidad de La Laguna (ULL), E-38205 La Laguna, Tenerife, Spain}
\altaffiltext{9}{Department of Astronomy, Yale University, 260 Whitney Avenue, New Haven, CT 06511, USA}

\begin{abstract}
The Hubble Frontier Fields (HFF) program combines the capabilities of
the \emph{Hubble Space Telescope} (\emph{HST}) with the gravitational
lensing of massive galaxy clusters to probe the distant Universe to an
unprecedented depth. Here, we present the results of the first combined
\emph{HST} and \emph{Spitzer} observations of the cluster Abell-2744. We combine the full near-infrared data with ancillary optical images to
search for gravitationally lensed high-redshift ($z \gtrsim 6$)
galaxies. We report the detection of 15 $I_{814}$-dropout candidates at $z \sim 6-7$ and one
$Y_{105}$-dropout at $z \sim 8$ in a total survey area of 1.43 arcmin$^{2}$ in the
source plane. The predictions of our lens model allow us to also
identify five multiply-imaged systems lying at redshifts between $z \sim 6$
and $z \sim 8$. Thanks to constraints from the mass distribution in
the cluster, we were able to estimate the effective survey volume
corrected for completeness and magnification effects. This was in turn used to
estimate the rest-frame ultraviolet luminosity
function (LF) at $z \sim 6-8$. Our LF results are generally in agreement with the most recent blank field estimates, confirming the feasibility of surveys through lensing clusters. Although based on a shallower observations than what will be achieved in the final dataset including the full ACS observations, the LF presented here goes down to \muv\ $\sim -18.5$, corresponding to 0.2L$^{\star}$ at $z \sim 7$, with one identified object at \muv\ $\sim -15$ thanks to the highly-magnified survey areas. This early study forecasts the power of using massive galaxy clusters as cosmic telescopes and its complementarity to blank fields.   

\end{abstract}


\section{Introduction} 

The discovery of the first galaxies to assemble in the Universe has long been one of the most exciting challenges of observational cosmology. The identification of these high-redshift sources in deep blank fields relies mostly on the photometric detection of the IGM absorption blueward \lya, or the Lyman break technique \citep{steidel96,giavalisco04}. Great progress has been made in characterizing the early galaxy population at $z \sim 6-7$ through the determination of their UV colors, stellar masses or ages \citep{on_Koekemoer_Stark_Bowler_2011,ollo_Gonzalez_Smit_et_al__2012, richard12} owing to unprecedented capabilities of the Wide Field Camera 3 (WFC3) onboard {\em HST}. Far from the several thousands of galaxies confirmed up to a redshift of 6.5, we are starting to spectroscopically confirm few galaxies at $z > 7$ \citep{vanzella11,schenker12,ono12,finkelstein13}. The decrease of the prevalence of \lya\ emitters at $z > 7$ might be a direct indication of the increase in the opacity of the intergalactic medium (IGM), which greatly attenuates the \lya\ line through resonant scattering \citep[e.g.,][]{stark10}. Recent studies also suggest that the \lya\ luminosity function decreases at $z > 7$ indicating an increase in the neutral gas fraction in the intergalactic medium \citep[IGM,][]{guchi_Ly_Nagao_Iye_et_al__2006,mada_Ota_Kashikawa_et_al__2010}. However, such conclusions are still prone to large uncertainties because of the small sample size, as near-infrared (NIR) observations of faint sources at $z > 7$ is extremely challenging. Of course, one key driver for these studies of early galaxies is to determine the sources responsible for re-ionization of the high redshift Universe. 

A complementary approach is to exploit gravitational lensing offered by massive galaxy clusters \citep{kneibnatarajan2011}, which magnifies the brightness of intrinsically faint sources. This has been successfully used to detect galaxies over a wide redshift range, taking advantage of the flux magnification \citep[e.g.][]{kneib2004,bouwens09,richard12,alavi13} and the higher spatial resolution for detailed, small-scale studies of high-redshift galaxies \citep[e.g.,][]{brammer12}. In this framework, the CLASH\footnote{\url{http://www-int.stsci.edu/~postman/CLASH/Home.html}} program \citep{postman12} yielded important results in a variety of fields using 25 lensing clusters, resulting in a significant progress in cluster mass modeling \citep[e.g.][]{zitrin12,bradley13,coe13, vanzella13}. Combining the exquisite capabilities of {\em HST} with the power of ``gravitational telescopes'', the new {\em HST} Frontier Fields\footnote{\url{http://www.stsci.edu/hst/campaigns/frontier-fields/}} (HFF) initiative is aiming to peer deeper into the distant Universe by initially devoting a total of 560 orbits to observe four clusters down to a magnitude limit of $\sim 29$. Over Cycles 21 and 22, the program will obtain deep optical and NIR imaging in seven filters using ACS and WFC3 for both the clusters and the parallel blank fields.  

 Here we present the first results of the NIR observations of the Abell-2744, where we search for high-redshift candidates at $z \sim 6-8$ behind the galaxy cluster. In Section \ref{sec:obs}, we present the observational dataset and reduction steps leading to the construction of the source catalog. The lensing model is described in Section \ref{sec:models}. The procedure used to select high-redshift candidates is detailed in Section \ref{sec:selection} together with the sources of contamination. In Section \ref{sec:lf} we present the luminosity distribution at $z \sim 6-8$ and compare to previous blank field results. We use a standard $\Lambda$CDM cosmology with $H_0=71$\ km s$^{-1}$\ Mpc$^{-1}$, $\Omega_{\Lambda}=0.73,$\ and $\Omega_{m}=0.27$. Magnitudes are in AB system.

\begin{figure*}[!htbp]
\begin{center}
\includegraphics[width=9cm]{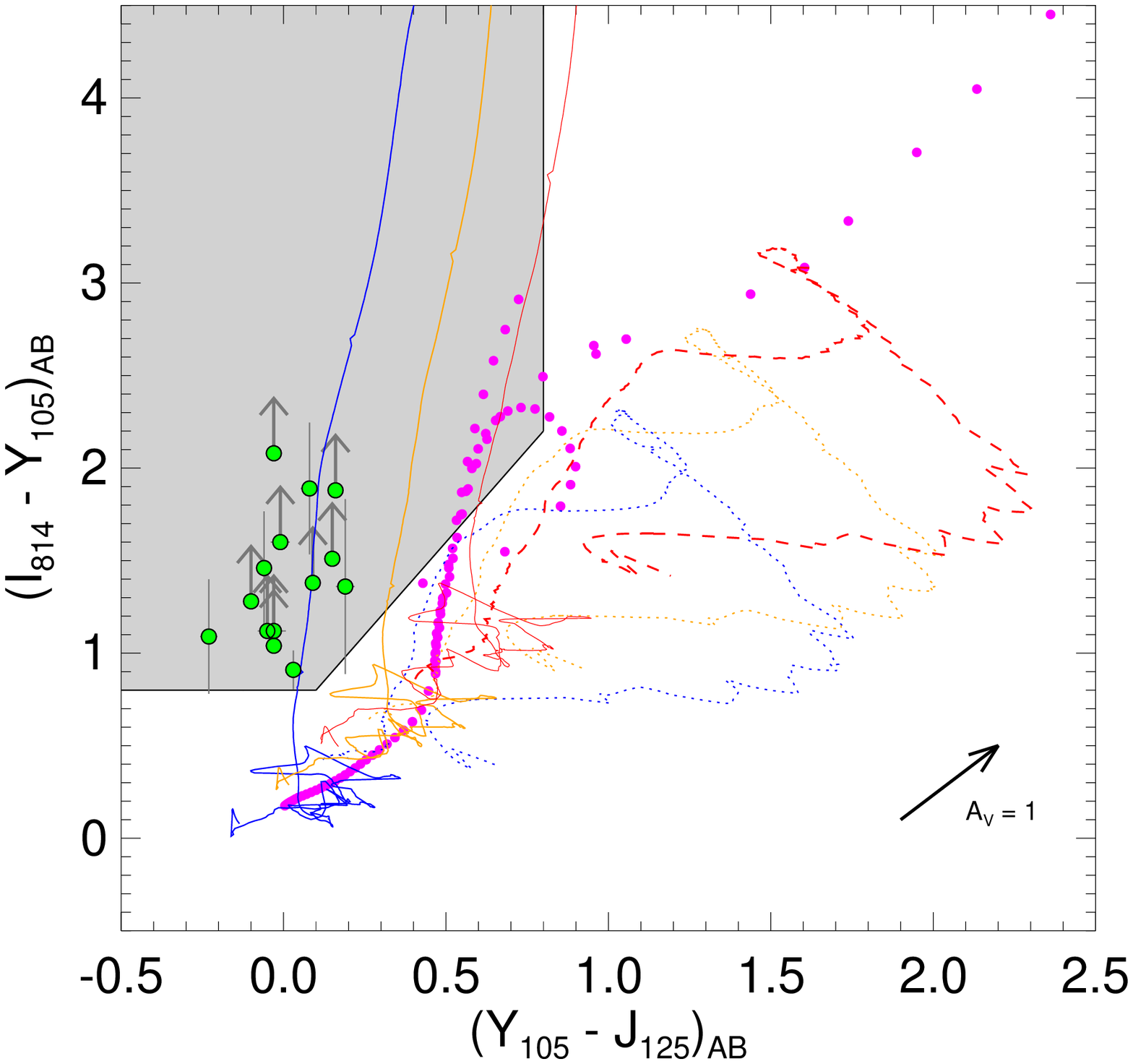}
\hspace{-0.3cm}
\includegraphics[width=9cm]{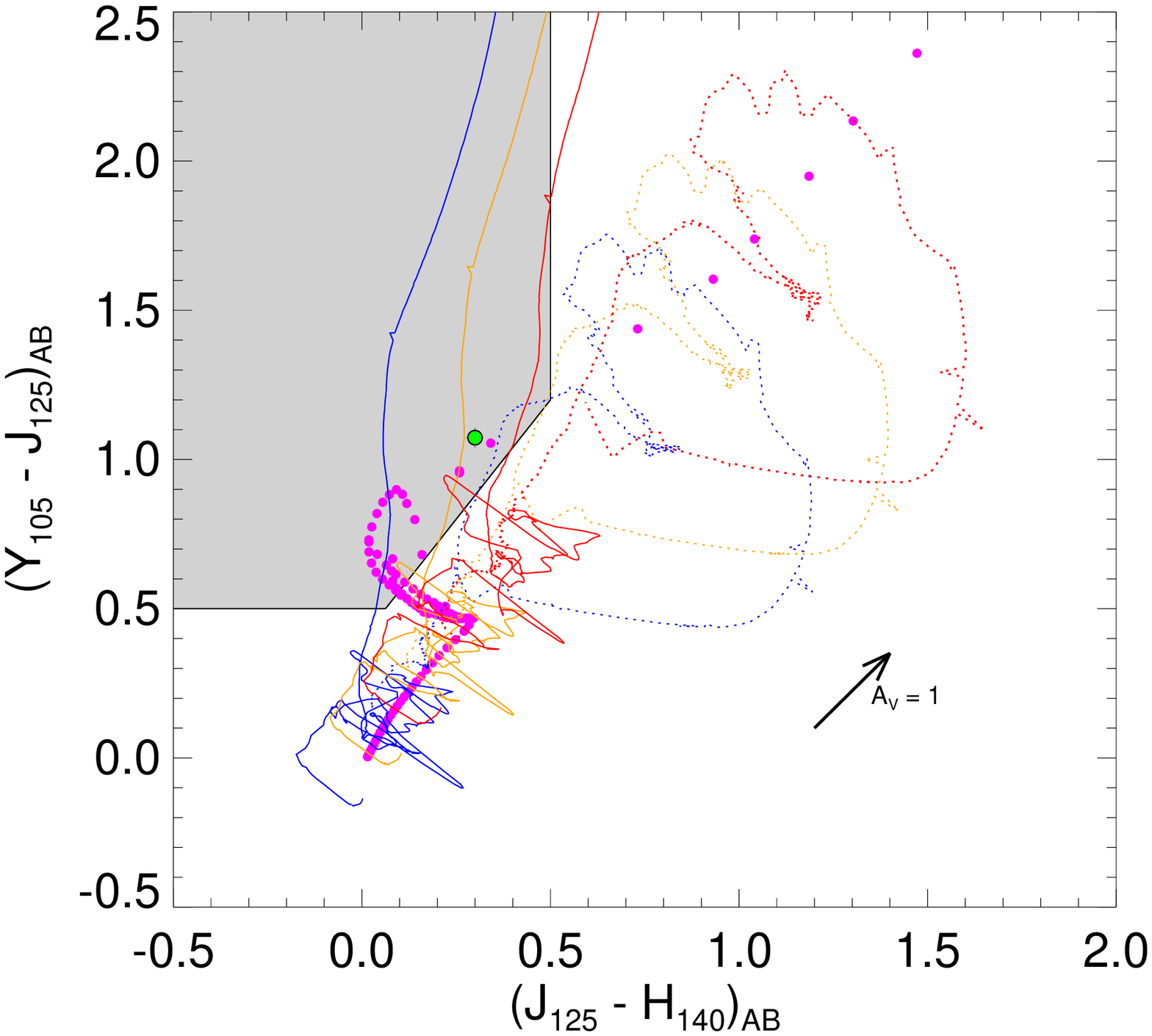}
\caption{\label{fig:color_selection}
Color-Color selection of $z \sim 6-7$ (left panel) and $z \sim 8$ (right panel) candidates. The dropout candidates are represented by green circles with 1-$\sigma$ uncertainties and the selection window by the shaded region. We also show the expected color tracks as a function of redshift for starburst galaxies (solid lines) and elliptical galaxies (dotted lines). The color code indicates an increase of attenuation from A$_{V} =0$ (in blue) to A$_{V}=2$ (in red). We used standard galaxy templates from of \citet{coleman80} and \citet{Storchi-Bergmann_Schmitt_1996}. The magenta points represent the colors of cool stars from \citet{chabrier00} catalog.}
\end{center}
\end{figure*}

\section{Observations and Data Reduction}
\label{sec:obs}

In this paper we combine existing imaging and spectroscopic data obtained in previous {\em HST} campaigns targeting A2477 cluster with new {\em HST} IR observations that were taken as part of the Frontier Fields program. The resulting pseudo-color image is shown in Figure \ref{fig:layout}.

\subsection{Previous {\em HST}  Data}

Optical observations were obtained with the Advanced Camera for Survey (ACS) onboard {\em HST} in cycle 17 (GO 11689, PI: Dupke) and used in the strong lensing analysis presented in \citet{merten_2011}. Images were taken with the Wide Field Camera (WFC) using three broadband filters F435W, F606W, and F814W. A summary of the dataset is presented in Table \ref{tab:obs}. For the basic reduction steps, we used the {\tt CALACS} package v2012.2 that include Charge Transfer Efficiency (CTE) corrections, which were not available in the publicly-released reductions. Then all exposures in each filter were median combined using {\tt Astrodrizzle} task in the new {STSDAS/Drizzlepac} package\footnote{\url{http://drizzlepac.stsci.edu}}. First, the different exposures were corrected for small misalignments using {\tt Tweakreg} before rejection of cosmic rays and correction for geometrical distortions . The final images have a pixel scale of 0.05\arcsec\ pix$^{-1}$ and reach a 5$-\sigma$ depth (for point sources in 0.4\arcsec\ aperture) of 27.6, 27.5, and 27.4 in F435W, F606W, and F814W, respectively. Depths were determined by measuring the median standard deviation from a hundred of apertures of 0.4\arcsec\ diameter positioned randomly on the sky.

\subsection{Hubble Frontier Fields Data}

The NIR observations of the first cluster in the HFF program (GO/DD 13495), using the Wide Field Camera 3 (WFC3), started on October 25th 2013. This includes imaging in four filters F105W, F125W, F140W, F160W, that achieves a total exposure time of 24, 12, 10 ,and 24 orbits, respectively. Here we use the full NIR observations summarized in Table \ref{tab:obs}. This is the first epoch observations of Abell 2744 that will be completed with ACS observations during the second epoch scheduled for June 2014. Basic reductions were once again performed using {\tt HSTCAL} and most recent calibration files. Individual frames were coadded using {\tt Astrodrizzle} after registration to the ACS reference image using {\tt Tweakreg} shift file. After an iterative process, we achieve an alignment accuracy of 0.1 pixel between WFC3 and ACS images. The final images have 0.13 \arcsec\ pixel size and a 5$-\sigma$ depth reached in the NIR filters is 28.6 (F105W), 28.5 (F125W), 28.5 (F140W), and 28.4 (F160W) respectively. 

\begin{table}
\caption{\label{tab:obs} Summary of {\em HST} observations}
\begin{tabular}{lccc}
 Instrument/Filter & \# Orbits\footnote{exposure time in ks for IRAC} & 5$\sigma$ Depth \footnote{3$\sigma$ depth in $\mu$Jy for IRAC data} & Obs Date \\ \hline 
  WFC3/F160W  & 24  & 28.4&Oct/Nov 2013\\
 WFC3/F140W  & 10  & 28.5&Oct/Nov 2013  \\
 WFC3/F125W  & 12  & 28.5&Oct/Nov 2013 \\
 WFC3/F105W  & 24  & 28.6&Oct/Nov 2013 \\
 ACS/F814W & 5  & 27.4 &Oct 2009\\ 
 ACS/F606W & 5 & 27.5&Oct 2009 \\ 
 ACS/F435W & 6  & 27.6 &Oct 2009 \\ 
 IRAC/3.6     & 90.9 & 0.139 & Sept 2013\\
 IRAC/4.5     & 90.9 & 0.225 & Sept 2013 \\ \hline
\vspace{0.3cm}
\end{tabular}
\end{table}

\begin{figure}[!htbp]

\centering
\includegraphics[width=9cm]{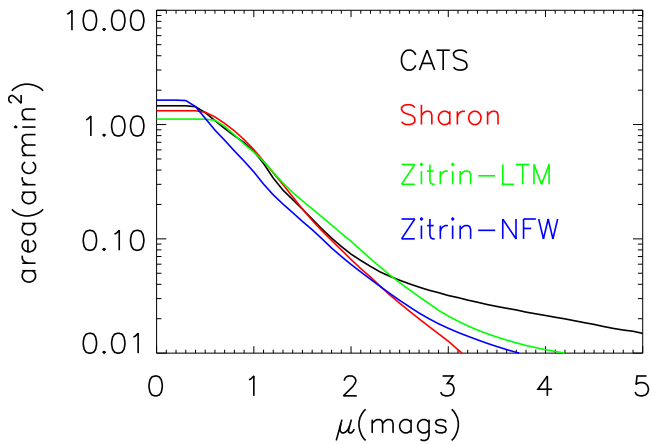}
\caption{\label{fig:surface} Survey area as a function of magnification in Abell 2744. The different curves represent the cumulative area at redshift $z =7$ probed for a given minimum magnification for different models. We use the mass models available on MAST\footnote{http://archive.stsci.edu/prepds/frontier/lensmodels/} that use a parametric approach (see Sect. \ref{sec:models} for details).}

\end{figure}

\begin{figure*}[tb]
\begin{center}
\includegraphics[width=15cm]{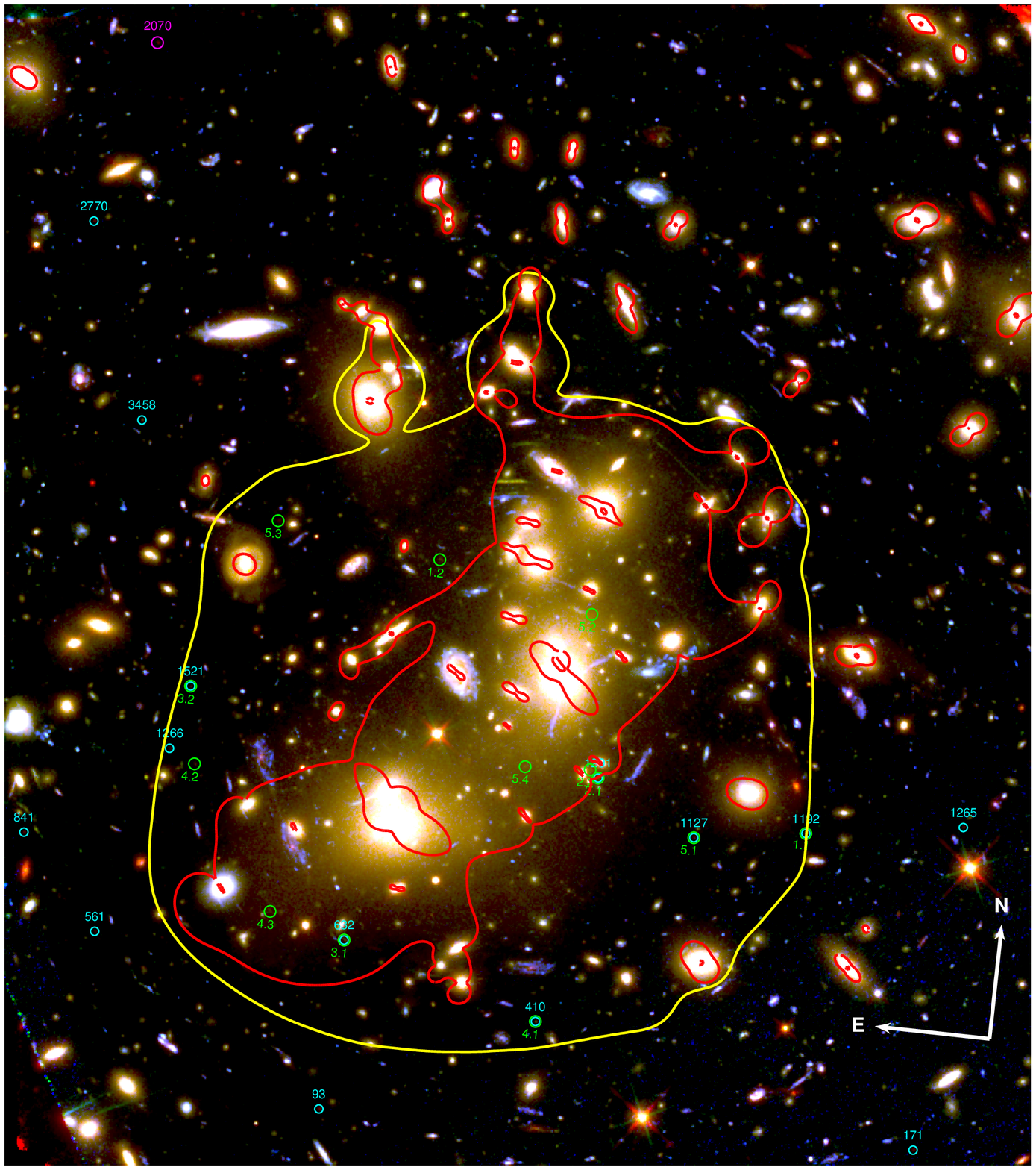}
\caption{\label{fig:layout}
The location of the drop-out images is superposed on the strong lensing model for Abell 2744 (Richard et al., in prep). The pseudo-color image is a combination of the F606W, F814W and a deep (four IR bands stack) WFC3/IR image. Overlaid in red are the critical lines for background sources at $z=7$. The position of the candidates at $z=6-7$ and $z\sim8$ are marked with cyan and magenta circles, respectively. The green circles show the positions of the multiple-image systems identified with the predictions from the mass model, while the yellow line is the predicted region for multiple images.}
\end{center}
\end{figure*}

\subsection{{\em Spitzer}  Data}

{\em Spitzer} imaging of Abell 2744 using Infrared Array Camera (IRAC) was obtained on September 2013 as part of the Frontier Field Spitzer program (PI : T. Soifer). The dataset we used here represents 50\% of the planned observations for the cluster. The total exposure time is 25 hours on sources in each of the 2 IRAC channels 3.6 and 4.5  microns. We used corrected Basic Calibrated Data (cBCD) images, that are provided by the Spitzer Science Center and automatically corrected by pipeline for various artifacts (such as muxbleed, muxstripe, and pulldown). The cBCD frames and associated mask and uncertainty images were processed, drizzled and combined into final mosaics using
the standard SSC reduction software MOPEX. The mosaic has a 2-$\sigma$ point source sensitivity (measured from the noise in a clean area) of 0.093 and 0.148 $\mu Jy$ in the 3.6 and 4.5 micron channel respectively.

\subsection{Photometric Catalogs}

We used a deep (F125W+F140W) image for source detection with SExtractor software \citep{Bertin_Arnouts_1996}. First, all images were PSF-matched to the F160W image with a PSF model derived using {\tt Tiny Tim} \citep{Krist_Hook_Stoehr_2011}  and to a pixel scale of 0.13\arcsec. Then we ran Sextractor in dual mode with the deep IR image as detection frame and the image in each filter to perform the photometry within the same aperture. The isophotal magnitude is adopted for the color, the total magnitude for the flux, and a local background calculation is adopted for the photometry. The final drizzling was performed with the inverse variance map to account for all the sources of uncertainties in the image. Then this weight map was transformed to an {\em rms} map including the correlated noise correction \citep{r_Hook_Levay_Lucas_et_al__2000} to derive flux uncertainties during source extraction.

\section{Cluster Mass Model}
\label{sec:models}

The SL mass model will be presented in details in a forthcoming publication (Richard et al. 2014, in prep.). We here outline the main points of the methodology. We follow our previous lensing work \citep[\emph{e.g.}][]{Richard2010,Limousin2012}, adopting a parametric mass model combining both large scale (cluster or group size) mass clumps and galaxy-scale mass clumps centred on each cluster galaxy (selected using a colour-magnitude diagram), using scaling laws based on their magnitude in order to assign a mass to each galaxy \citep{Limousin2007}. The number of large scale mass clumps is driven by the goal to reproduce as accurately as possible the location of the multiply-imaged systems used as constraints.

Starting with an earlier model by \citet{Merten2011}, we looked for multiply imaged system within the ACS field of view. We finally identified 52 images coming from 17 multiply-imaged background sources. Five of them (corresponding to two background sources) are located close to sub-clumps (group scale) located to the North and North West limits of the ACS coverage (outside the WFC3 field of view). We targeted the core of this cluster with FORS2 on the ESO {\em Very Large Telescope} ({\em VLT}) and measured spectroscopic redshifts for two systems, allowing to reliably calibrate the mass model. The optimization is performed in the image plane, using the \textsc{Lenstool} software \citep{Jullo07}. We found that a five-component mass model is able to reproduce the observational constraints with an RMS in the image plane equal to 0.7$\arcsec$. Three large scale mass clumps are located in the core of the cluster and are associated with the brightest galaxies. The other two large scale mass clumps are associated with the two North and North West sub-clumps.
On top of these large scale mass clumps, perturbations associated with cluster members are considered. From this mass model, we generate the critical lines and the multiple-image region displayed in Fig. \ref{fig:layout} and compute the magnification map.

\section{Candidate Selection}
\label{sec:selection}

For the selection of high-redshift candidates, we adopted the commonly used dropout criteria based on the Lyman break technique \citep{steidel96,giavalisco04}. For $I_{814}$ dropouts we adopted the following criteria \citep[see for example][]{oesch10}:

\begin{align}
\label{eq:criteria7}
 (I_{814} {-} Y_{105})   &>  0.8  \notag \\
(I_{814} {-} Y_{105})   &>  0.6 + 2(Y_{105} {-} J_{125})\\
(Y_{105} {-} J_{125})  &< 0.8 \notag
\end{align}

In order to more efficiently reject low$-z$ interlopers the second criterion is more stringent than what has been used previously to identify similar sources (see Fig. \ref{fig:color_selection}). We also require that the candidates are detected in $Y_{105}$ and $J_{125}$ bands with a minimum of $5-\sigma$ significance while they remain undetected in the optical $B_{435}$ and $V_{606}$ bands at less than $2-\sigma$. Objects that are not detected in the $I_{814}$ filter are assigned a $3-\sigma$ lower limit for their continuum break. To identify the $Y_{105}$ dropouts we applied the following color selection:  

\begin{align}
\label{eq:criteria8}
(Y_{105} {-} J_{125})  &>  0.5 \notag \\
(Y_{105} {-} J_{125})  &>  0.4 + 1.6(J_{125} {-} H_{140})\\
(J_{125} {-} H_{140})  &< 0.5 \notag
\end{align}

Similarly, we require a $5\sigma$ detection in the bands redward of the break and no detection in all the bands blueward the break. The selection procedure is illustrated in Fig. \ref{fig:color_selection}, where the left and right panels show the $z \sim 6-7$ and $z \sim 8$ selections, respectively. Objects satisfying our dropout selection (shaded box) are represented by green circles with $1-\sigma$ uncertainties. We also plot the color tracks of different galaxy types as a function of redshift and for three values of attenuation. We used standard libraries of \citet{coleman80} to calculate the tracks of elliptical galaxies (dotted lines) and \citet{Storchi-Bergmann_Schmitt_1996} for star-forming galaxies (solid lines). The blue, orange and red colors represent an attenuation of $A_{V}$=0, 1, and 2, respectively. All objects are visually inspected to remove spurious detections or contaminated photometry from close objects. Our final sample consists of 15 candidates at $z\sim 7$ and one candidate at $ z \sim 8$. Object 3284 is very uncertain due to its point-like morphology. Therefore we decided to not include this candidate in the $z \sim 7$ luminosity function determination of Sect. \ref{sec:lf}. The $Y_{105}$-dropout shows a flux excess in the 4.5 $\mu m$ IRAC band while it is undetected in the 3.5 $\mu m$ channel. This is possibly due to the contribution of the redshifted [O{\sc III}] + H$\beta$ emission lines. A detailed analysis of the properties of this candidate and discussion of its high-redshift solution are presented in \citep{laporte14}. Gravitational lensing also provides higher angular resolution allowing the study of spatially resolved star formation and internal structure of distant galaxies \citep{frye12,brammer12}. The HFF observations will extend this study to the most magnified high-redshift galaxies. Candidate 561 for instance shows an arc-like extended morphology. The source reconstruction of such objects will enable us to analyze the clumps and morphology of the building blocks of present day galaxies and compare them to other results at intermediate redshifts \citep{zitrin11,bradley12}. Because the optical ACS data do not match the depth of the new IR observations, we restricted the search to relatively bright candidates. Specifically, in the case of objects undetected in the $I_{814}$ filter, we measured the 2-$\sigma$ limit in the same aperture as in the IR. Then we retained only candidates that satisfy the selection criteria \ref{eq:criteria7}, replacing the $I_{814}$ magnitude by its limit. For both $I$- and $Y$-dropouts, to ensure non-detections in the optical bands blueward the break, the detection limit in the combined optical image was about one magnitude deeper than the object magnitude in the detection band. Consequently, the faintest object in our catalogue was about $J_{125} \sim$ 27.5 mag. The identification of fainter objects will be possible with upcoming deep ACS observations that will be taken as part of the same HFF program.

\begin{figure}[!htbp]
\vspace{0.2cm}
\centering
\includegraphics[width=8.8cm]{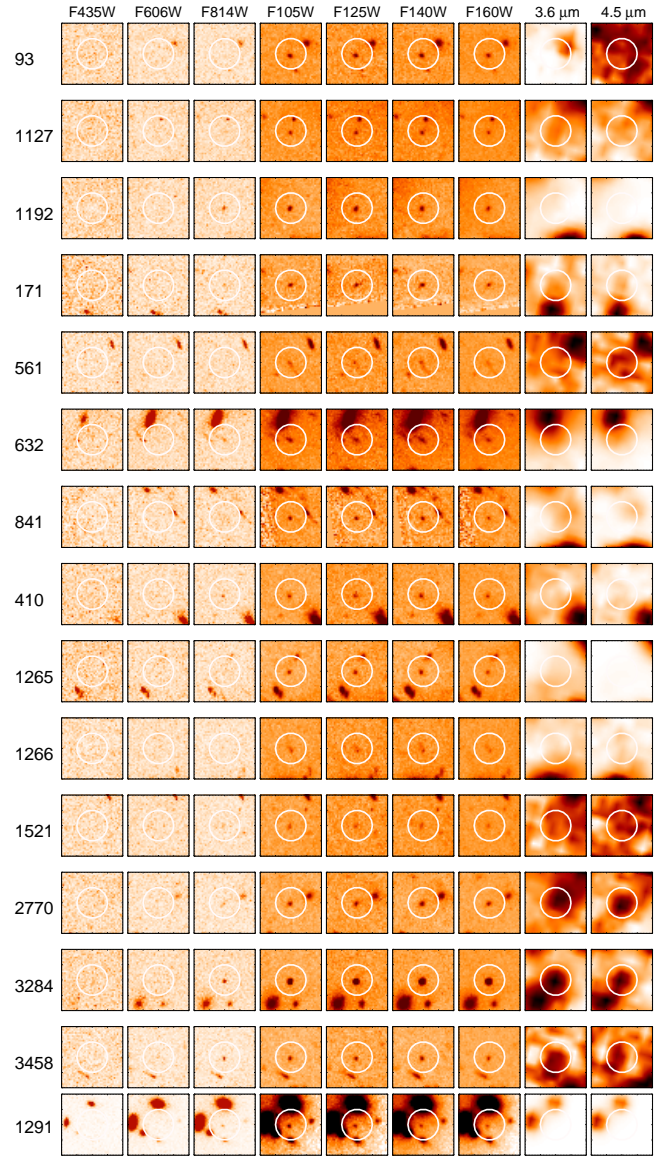}
\caption{\label{fig:z7_cutout}
Postage stamps of the $z \sim 6-7$ candidates in the ACS $F435W, F606W, F814W$, WFC3 $F105W, F125W, F140W$, $F160W$ and IRAC 3.6 $\mu m$, 4.5 $\mu m$ bands. The size of each cutout is about 5\arcsec and the white circle denotes the source position. Sources show a strong $I_{814} - Y_{105}$ and remain undetected at a $2-\sigma$ level in the $B_{435}$ and $V_{606}$ bands.
}
\vspace{0.2cm}
\end{figure}

\begin{figure}[!htbp]
\vspace{0.2cm}
\centering
\includegraphics[width=8.8cm]{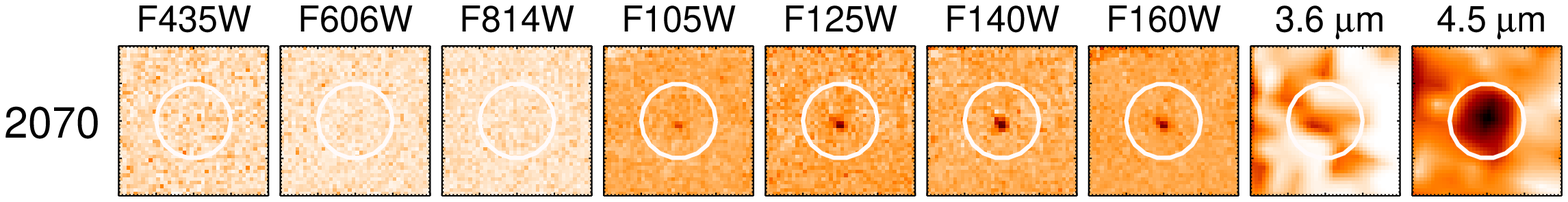}
\caption{\label{fig:z8_cutout}
same as Fig. \ref{fig:z7_cutout} for a $z \sim 8$ candidate. The source has a strong $Y_{105} - J_{125}$ break with no detection at a $2-\sigma$ level in the $B_{435}, V_{606}$, and $I_{814}$ bands. It is detected in all WFC3 filters redward the break with at a minimum of $5-\sigma$ significance.}
\vspace{0.2cm}
\end{figure}

\subsection{Contamination}

Given the the stringent selection criteria we used here, we expect very low contamination rate from spurious sources or low-redshift interlopers. As we have seen before, the IR observations are much deeper than the faintest candidate in our sample. This ensures that all candidates were detected in several filters with a very high significance, which basically excludes spurious detections.

Since the infrared data have been taken much later than the ACS data, transient sources such as supernovae that exploded recently, can be detected only in the infrared satisfying the continuum break criterion. We first compared the four epochs of IR observations spanning a period of five weeks and verified the consistency between the photometry in all the bands. furthermore, such sources would have appeared as point sources. The inspection of the high-resolution images of the HST shows that the candidates are all spatially resolved.

Possible sources of contamination also include low-redshift galaxies that enter the selection space. This can occur when strong nebular emission lines contribute to the total flux in one filter, mimicking a continuum break \citep{bert_Bridge_Bunker_et_al__2011}. However, the use of multiple and contiguous filters greatly mitigates the contamination from such sources, since the emission line should be isolated and the flux enhancement restricted to only one filter (or two in the case of a combination of lines such as $\ha$ and [OIII]). Low-$z$ interlopers with unusually high reddening or extremely old stellar population, although rare, could also be selected mistakenly as high-$z$ dropouts \citep{Pello_Schaerer_Le_Borgne_2012}. The colors of most of our candidates suggest a rather blue or flat continuum incompatible with the existence of such a red population. 

Finally, cool dwarf stars can have similar colors to high-redshift galaxies. As we can see in Fig. \ref{fig:color_selection}, brown dwarfs color track is more a concern for $z \sim 8$ galaxy selection since it comes close to our candidate. However, we stress again that stars can be visually identified as point-like sources from their light profiles or the stellarity parameter of SExtractor (i.e. stellarity $> 0.6$).

\subsection{Photometric Redshifts}

Additionally, we computed photometric redshifts for our candidates by fitting the photometric data with spectral energy distribution (SED) templates using the Hyperz software \citep{bolzonella00}. We used standard stellar population libraries of \citet{Bruzual_Charlot_2003} and a \citet{Salpeter_1955} initial mass function (IMF). We find that the probability distributions of the photometric redshifts all indicate a high-$z$ solution in agreement with the color-color selection. The photometric and color information of the candidates are presented in Table \ref{tab:photometry}. We also include in the same table the best-fit redshift given by the probability distribution function for each galaxy, together with the magnification factors. Figures \ref{fig:z7_cutout} and \ref{fig:z8_cutout} show the image cutouts of the $z \sim 6-7$ and 8 candidates in the ACS, WFC3, and IRAC filters.

\subsection{Multiple Images}

Strong lensing creates multiple images of the same background galaxy, whose locations can be predicted by our lens model (Richard et al., in prep). For each object selected by our color-color selection, we visually search the field for additional counter-images according to their photometric redshift, color and position. We also looked for multiply-imaged systems that did not satisfy our selection criteria, mainly because of their low signal to noise ratio or contamination from cluster members. In Table \ref{tab:photometry} we keep only the brightest image amongst the multiples of the same object. The full catalog of multiple images is shown in Table \ref{tab:multiple}. We identified two double-image, two triple-image, and one quadruple-image system between $6 \lesssim z \lesssim 8$, which are also presented in Figure \ref{fig:layout} with their respective identifying indices. The counter-images 1.2, 2.1, 5.2, 5.4 were too severely affected by bright nearby objects and/or had a low signal-to-noise ratio to be included in our initial dropout selection. Their photometric redshift was however consistent with their dropout counterpart. In order to further assess the robustness of these multiply imaged candidates, we included each of them separately in the mass model as new constraints. We found that all candidates fit into the mass model without significant deviation, their individual RMS being of the order of the total RMS. Since parametric strong lensing mass modeling is very sensitive to redshift misidentification, we consider this test as a further confirmation of the robustness of these systems.

Some of these counter-images will likely be confirmed with the forthcoming ACS observations of Abell 2744 which will provide deeper optical images allowing to test their multiplicity and thus corroborating the high redshift nature of these candidates \citep[e.g.,][]{ellis2001,kneib2004} as well as the robustness of our lens modeling procedure.

\section{The UV Luminosity Function at $z=7-8$}
\label{sec:lf}

\begin{figure}[tb]
\begin{center}
\includegraphics[width=7.5cm]{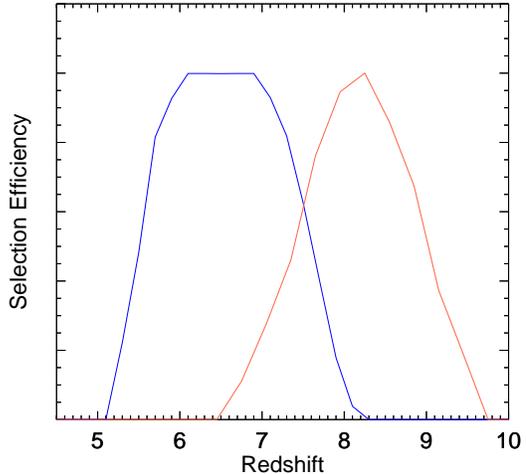}
\caption{\label{fig:zfunc} Redshift selection function showing the relative efficiency in selecting $I_{814}$ (blue curve) and $Y_{105}$ (red) galaxies. This is determined using the simulations of color-color selection and completeness described in Sect. \ref{sec:volume}.}
\end{center}
\end{figure}

We now turn to the estimate of the rest-frame UV luminosity function (LF) at redshift $z \sim 6-7$ and 8 using the two galaxy samples assembled above. We first determine the absolute magnitude of each source in the $J_{125}$ band at a mean redshift of $z \sim 6-7$ and in the $H_{140}$ filter at $z \sim 8$. The observed values need to be corrected for the gravitational lensing magnification.

\subsection{Effective Volume}
\label{sec:volume}

While amplifying the intrinsic flux of a given source, strong lensing also distorts and stretches the source plane volume where it lies. The resultant drawback is that a higher-magnification region will necessarily probe a smaller comoving volume. Therefore, we need to account for these two effects in our LF estimates. The LF is given by

\begin{eqnarray}
\phi(M)dM = \frac{N_{i}}{V_{eff}(M_{i})}
\end{eqnarray}

where $N_{i}$ is the number of galaxies in the $i$th bin and $V_{eff}$ is the associated effective survey volume.

\begin{figure*}[!htbp]
\begin{center}
\hspace{-0.5cm}
\includegraphics[width=9.5cm]{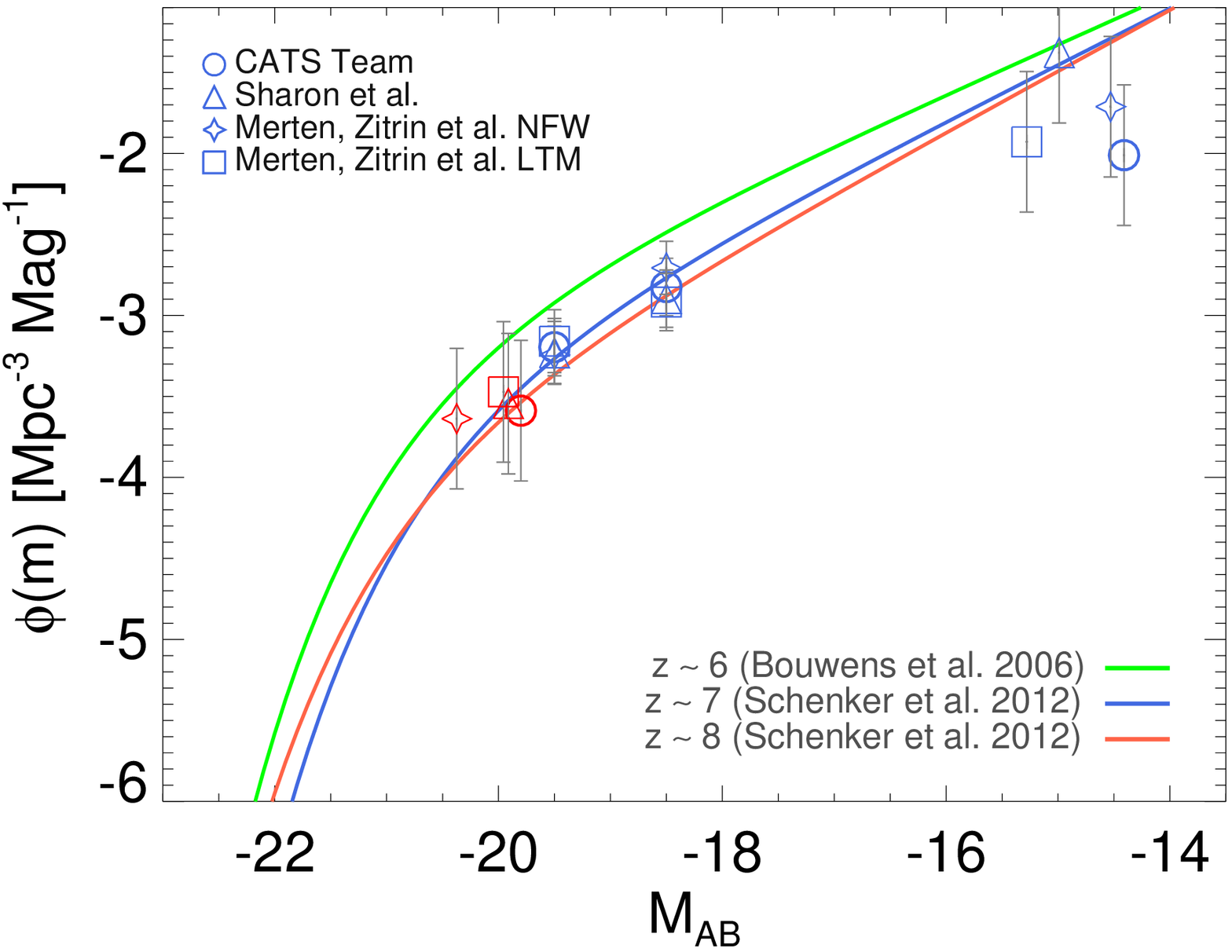}
\hspace{-0.6cm}
\includegraphics[width=9.4cm]{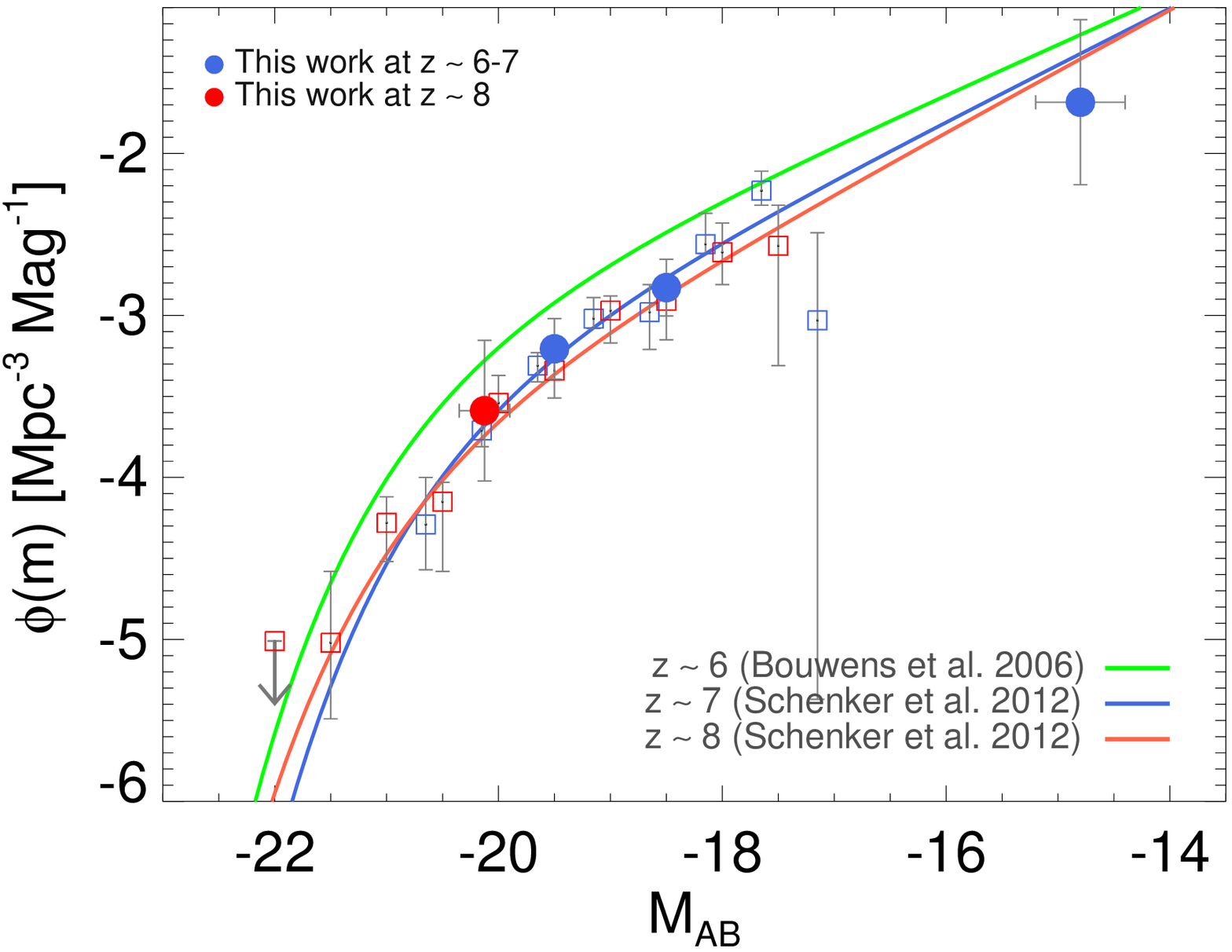}
\hspace{-1.1cm}
\caption{\label{fig:lf} Rest-frame UV luminosity function at $z \sim 6-8$ based on our $I_{814}$ and $Y{105}$ dropout samples. {\it Left:} UV LF determination using four different lensing models repented by different symbols (cf. legend). The blue symbols represent the LF determination at $z\sim 6-7$ compared to previous results in the blank fields at $z \sim 6$ \citep[green line,][]{bouwens06} and $z \sim 7$ \citep[blue line,][]{schenker12}. The red symbols are the estimate at $z \sim 8$ compared to the HUDF12 determination at the same redshift \citep[red line,][]{schenker12}. Error bars correspond to $1-\sigma$ uncertainties. {Right:} The UV LF averaged over the four models. The blue circles are the LF at $z \sim 6-7$ compared to the HUDF12 resutts at $z \sim 7$ (blue open squares) and $z \sim 8$ (red open squares).}
\vspace{0.9cm}
\end{center}
\end{figure*}

Our effective volume is also determined by the shape of the redshift selection function and the incompleteness. We generated starburst templates from \citet{Storchi-Bergmann_Schmitt_1996} library, which were shifted to the desired redshifts $5 < z < 10$, then we applied attenuation of A$_{V}=0-1$, and calculated synthetic fluxes using the {\em HST} filter throughputs. Then we created artificial galaxies using {\tt ARTDATA} package in IRAF, exploring the parameter space of observed magnitude, color, shape, and position in the image. We generated 10,000 galaxies randomly added to the actual optical and IR images, using 100 objects at a time. We applied a distribution of absolute magnitudes between -23 and -14 and applied the colors derived from the spectral templates. We used a log-normal distribution of half light radii derived from the observed sizes of spectroscopically confirmed $z \sim 4$ LBGs of \citet{vanzella09}. We applied a factor of (1+$z$)$^{-1}$ to account for redshift-evolution in size \citep[see][]{grazian11}. The sizes are computed in the source plane and the galaxy positions chosen randomly in the image plane. Then we used Lenstool to simulate the size and flux of the magnified galaxies in the image plane using our A2744 mass model. In the completeness calculation, we keep only the brightest image of each system.

We applied our source extraction and selection criteria \ref{eq:criteria7} and \ref{eq:criteria8} on the final images and compared to the input object catalogs. Given our stringent selection criteria, the incompleteness function is dominated by the contamination by bright sources in the crowded field. We incorporate this incompleteness correction in the selection function that goes into the effective volume calculation.

The effective survey volume for each magnitude bin is calculated according to the following equation
 
\begin{eqnarray}
V_{eff} = \int_{0}^{\infty}  \int_{\Omega > \Omega_{min}} \frac{dV_{com}}{dz}~ f(z,m,\mu) ~d\Omega(\mu,z) ~dz
\end{eqnarray}

where $\Omega_{min}$ is the source plane area with a minimum magnification $\mu_{min}$ required to detect a galaxy with an apparent magnitude $m$. $f(z,m,\mu)$ is the completeness function including the redshift selection function, and $d\Omega(\mu)$ is the area element in the source plane as a function of the magnification factor.

Several mass models for the HFF clusters are accessible through the STScI website including ours. In order to estimate the uncertainties in the magnification maps and the differences between the models, we derived the luminosity function using each model. Among the other five available models, only three offer the Kappa and Gamma maps that we integrate to derive the effective volume. These are the models of Sharon et al., Zitrin NFW, and Zitrin LTM. In addition to the effective surface that shows variations from one model to another, we also computed the amplification factor for our candidates using the four models.

\subsection{UV LF Results}

The results are shown in Fig. \ref{fig:lf} where our luminosity distribution is compared with previous results at $z \sim 6$ to 8. At redshifts $z =6-7$ we use two bins in magnitude (-18.5,-19.5) plus one object around M$_{abs} \sim -15$, whereas we only have one candidate at $z \sim 8$. We stress here that object 1291 is close to the $z$=7 critical line where the uncertainties become very high. Therefore, the amplification, hence the intrinsic magnitude of this candidate, must be taken with caution.

The left panel of Fig. \ref{fig:lf} illustrates how the uncertainties in the different lensing models affect the luminosity function determination. The four models yield a very similar luminosity function, although at high amplification (see Table \ref{tab:multiple}) the differences become significant (at M$_{abs} \sim -15$). The uncertainties on the UV LF determination also include the cosmic variance and poisson errors. We estimated the 1-$\sigma$ fractional uncertainties of the galaxy number counts using \citep{trenti08} cosmic variance tool with the selection function and survey volume as inputs. Because of the small number statistics, the fractional poisson errors (1/$\sqrt{N}$) are comparable to the uncertainties due to cosmic variance, which account to up to $\sim 30\%$.

We show in the right panel of Fig. \ref{fig:lf} the averaged LF constructed from all the models. Overall, our derived LF is in good agreement with blank field results from the Hubble Ultra Deep Field \citep[HUDF,][]{bouwens06} and HUDF12 \citep{schenker12}. Most notably, we report here the faintest galaxy discovered at these redshifts, even though the uncertainties associated to its proximity to the critical line (where amplification is virtually infinite) are very large. The nature of this candidate will be better constrained with the advent of deep ACS data as part of the second epoch of the HFF observations. In the meantime, our results show that strong lensing will enable us to probe the distant Universe at very faint magnitudes, comparable to intermediate redshift results, shedding light on the potential sources of cosmic reionization.

\section{Conclusion}
\label{sec:conclusion}

The first HFF result we report here is important in regards to unraveling the properties and distribution of high-redshift galaxies since it confirms previous survey results derived from completely independent fields and therefore unequivocally demonstrates the feasibility of surveys using gravitational lensing fields. The agreement with previous blank field determinations of the LF also points to the robustness of our mass models and hence our modeling procedure. Despite the current limited depth of the observations (soon to be rectified once the HFF/ACS observations are completed), our LF robustly reaches an intrinsic magnitude of M$_{abs} \sim -18.5$ at $z \sim 6-7$, which corresponds to about 0.2$L^{\star}_{z=7}$, and extends down to M$_{abs} \sim -15$ with a highly amplified object (a factor of 30 to 70). Albeit with large uncertainties, this result bolsters the advantage gained from gravitational magnification.

With the completed observations of the proposed six lensing clusters, the HFF program will probe the high-redshift Universe to unprecedented depths, about two magnitudes deeper than typical blank field surveys, with the aid and enhancement provided by these  ``cosmic telescopes''. We have shown here the feasibility and the effectiveness of such studies and the robustness of our cluster mass  models, which will gain even more accuracy as new candidates and ever more multiply-imaged systems are discovered, identified and confirmed. The LF of these highest redshift galaxies is a key signature and determinant of the sources responsible for re-ionization of the Universe. Unmasking these sources with the help of the additional magnifying power offered by cluster-lenses looks not only promising but also feasible as presented here with the first HFF results.

\acknowledgments

HA and JPK are supported by the European Research Council (ERC) advanced grant ``Light on the Dark'' (LIDA).
PN acknowledges support from NSF theory grant AST-1044455 and a theory grant from Space Telescope Science Institute HST-AR1214401.A. ML acknowledges support from CNRS.

\begin{table*}
\centering
\caption{\label{tab:photometry} Photometric and color measurements for the $z \sim 6-7$ dropouts}
\begin{tabular}{lccccccc}
\hline
Target & R.A. (J2000) & Dec (J2000) & $I_{814}-Y_{105}$ & $Y_{105}-J_{125}$ & $J_{125}\footnote{total magnitude}$  & Magnification\footnote{this is the flux amplification factor} & Photo$-z$  \\ \hline    
93    & 3.593807 &  -30.415442  &   $>$ 2.08    &   -0.03  $\pm$  0.02   &  26.39 $\pm$   0.01   & 3.42 $\pm$ 0.19 &	6.8  \\
171   & 3.570648 &  -30.414662  &   1.89   $\pm$  0.31   &  0.08  $\pm$  0.02   &  26.49 $\pm$   0.03   & 1.57 $\pm$ 0.03 & 6.3	  \\
410  & 3.585800 & -30.411740    &	 1.04$\pm$    0.15   &  -0.03  $\pm$  0.02   &  27.06  $\pm$  0.02  & 3.69   $\pm$ 0.18 & 5.8 \\
561   & 3.603225 &  -30.410330  &   $>$ 1.28  &    -0.10 $\pm$   0.03  &   27.04 $\pm$    0.03   &3.75 $\pm$ 0.19  & 7.5 \\
632   & 3.593541 &  -30.409719  &   1.46   $\pm$  0.16   &  -0.06   $\pm$ 0.02	 & 26.57  $\pm$  0.02  & 6.28 $\pm$ 0.56 &  5.9	  \\
841   & 3.606378 &  -30.407279  &   $>1.60 $    &  -0.01   $\pm$ 0.03	 & 26.73  $\pm$  0.02  &  2.27 $\pm$ 0.07 &  6.4	  \\
1127  & 3.580442 &  -30.405039  &   $>1.38 $  &  0.09   $\pm$ 0.03	&  26.73  $\pm$  0.02  &4.71 $\pm$ 0.36  &  6.4	   \\
1192  & 3.576128 &  -30.404494  &   1.23   $\pm$ 0.10	&  0.06   $\pm$ 0.01	&  26.46  $\pm$  0.01  & 2.92 $\pm$ 0.14 &  6.1	    \\
1265  & 3.570062 &  -30.403720  &   $> 1.51 $    &   0.15   $\pm$ 0.02	&  27.01  $\pm$  0.03  & 1.95 $\pm$ 0.05 &  7.0	   \\
1266  & 3.601099 &  -30.403956  &   $>1.36  $ &  0.19    $\pm$ 0.03	 & 26.94  $\pm$  0.02  &  3.51 $\pm$ 0.22 &  6.5	   \\
1291 & 3.584396 & -30.403395   &	 1.12$\pm$    0.17   &  -0.05  $\pm$  0.03   &  27.63  $\pm$  0.03  &  72.75  $\pm$ 9.75 &  6.0  \\
1521  & 3.600541 &  -30.401804  &   1.12   $\pm$ 0.22	&  -0.03   $\pm$ 0.04	&  27.50  $\pm$  0.05  & 3.04 $\pm$ 0.16 & 5.9	   \\
2770  & 3.606230 &  -30.386646  &   1.88   $\pm$0.15   &  0.16     $\pm$0.02	& 26.34   $\pm$ 0.03      &1.52 $\pm$ 0.03  &   5.5 \\
3284  & 3.576655 &  -30.391364  &   2.59   $\pm$0.07   &  0.22     $\pm$0.00	& 24.05   $\pm$ 0.00	  &5.82 $\pm$ 0.55  &  6.5   \\
3458  & 3.603540 &  -30.393106  &   0.91   $\pm$0.10   &  0.03    $\pm$0.02   &  26.68   $\pm$ 0.02	  & 1.77 $\pm$ 0.05 &  6.0   \\
\hline
Target & R.A. (J2000) & Dec (J2000) & $Y_{105}-J_{125}$ & $J_{125}-H_{140}$ & $H_{140}$    &  & \\   
\hline
2070 &  3.604522 &   -30.380463 &  1.07 $\pm$	0.03  &   0.30  $\pm$  0.02  &   26.22  $\pm$  0.01   & 1.47 $\pm$ 0.02 &  8.35  \\ \hline
\vspace{0.3cm}
\end{tabular}
\end{table*}

\begin{table*}
\centering
\caption{\label{tab:multiple} Multiple-image systems at $z \sim 6-8$}
\begin{tabular}{lcccccc}
\hline
Image & Target & R.A. (J2000) & Dec (J2000) & $I_{814}-Y_{105}$ & $Y_{105}-J_{125}$ & $J_{125}$\footnote{total magnitude}   \\ \hline 
1.1 & 1192 & 3.576128 & -30.404494&	 1.23$\pm$    0.10   &   0.06 $\pm$   0.01  &   26.46 $\pm$  0.02   \\
1.2 & 3822 & 3.591436 & -30.396687&	 1.66$\pm$    0.22   &  -0.07  $\pm$  0.03   &  27.34  $\pm$  0.03      \\
2.1 & 1291 & 3.584396 & -30.403395&	 1.12$\pm$    0.17   &  -0.05  $\pm$  0.03   &  27.83  $\pm$  0.09      \\
2.2 & 1291b& 3.584741 & -30.403147&        ...               &  ...                  & $>$ 27.8                     \\
3.1 & 632  & 3.593541 & -30.409719&	 1.46$\pm$    0.16   &  -0.06   $\pm$ 0.02    & 26.57   $\pm$ 0.02       \\
3.2 & 1521 & 3.600541 & -30.401804&	 1.12$\pm$    0.22   &  -0.03  $\pm$  0.04   &  27.64  $\pm$  0.03      \\
4.1 & 410  & 3.585800 & -30.411740&	 1.04$\pm$    0.15   &  -0.03  $\pm$  0.02   &  27.15  $\pm$  0.03      \\
4.2 & 1180 & 3.600068 & -30.404382&	 1.47$\pm$    0.26   &  0.23   $\pm$ 0.04    & 27.21   $\pm$ 0.03      \\
4.3 & 673  & 3.596525 & -30.409031&	 1.69$\pm$    0.43   &  0.33  $\pm$  0.07   &  27.32  $\pm$  0.04     \\
5.1 & 1127 & 3.580442 & -30.405039&	 1.89$\pm$    0.36   &  0.09  $\pm$  0.03   &  26.75  $\pm$  0.02     \\
5.2 & 1956 & 3.585321 & -30.397960&	$>$ 0.36      &   0.09  $\pm$  0.03   &  27.18  $\pm$  0.02     \\ 
5.3 & 3761 & 3.597850 & -30.395970&	$>$ 0.23     &   0.36  $\pm$  0.04   &  26.74  $\pm$  0.04     \\
5.4 & 1327 & 3.587268 & -30.403279&	$>$ 0.18      &   0.12  $\pm$  0.0.04   &  27.48  $\pm$  0.05      \\ \hline
\vspace{0.3cm}
\end{tabular}
\end{table*}

\bibliographystyle{apj}
\bibliography{references.bib}

\end{document}